# Tunneling magnetoresistance enhancement by symmetrization in spin-orbit torque magnetic tunnel junction

Jiaqi Zhou,[1,2] Weisheng Zhao,[1,*] Kaihua Cao,[1] Shouzhong Peng,[1] Zilu Wang,[1] and Arnaud Bournel[2]

[1]*Fert Beijing Institute, BDBC, School of Electronic and Information Engineering, Beihang University, Beijing 100191, China*

[2]*Centre de Nanosciences et de Nanotechnologies, CNRS, Université Paris-Sud, Université Paris-Saclay, Orsay 91405, France*

(Dated: May 11, 2018)

## Abstract

Heavy metals with strong spin-orbit coupling (SOC) have been employed to generate spin current to control the magnetization dynamics by spin-orbit torque (SOT). Magnetic tunnel junction based on SOT (SOT-MTJ) is a promising application with efficient writing operation. Unfortunately, SOT-MTJ faces the low tunneling magnetoresistance (TMR) problem. In this work, we present an *ab initio* calculation on the TMR in SOT-MTJ. It is demonstrated that TMR would be enhanced by SOT-MTJ symmetry structure. The symmetrization induces interfacial resonant states (IRSs). When IRSs match identical resonances at the opposite barrier interface, resonant tunneling occurs in SOT-MTJ, which significantly contributes to the conductance in parallel configuration and improves TMR. We demonstrate the occurence of resonant tunneling by transmission spectra, density of scattering states and differential density of states. We also point out that the thickness of heavy metal has limited influence on TMR. This work would benefit the TMR optimization in SOT-MTJ, as well as the SOT spintronics device.

## INTRODUCTION

The spin-orbit torque (SOT) originating from spin Hall effect (SHE) donates a novel approach to control the magnetization dynamics of the ferromagnet (FM) layers[1–4]. Due to the SHE, spin current induced by current in bottom heavy metal (HM) impose spin torques to the FM film, and switch the magnetic orientation with low switching current density[5–9]. Several HMs with large SHE have been used as the bottom HM, including W, Ta, Pt, Hf[10–12]. Besides, spin Hall angle as large as 0.3 has been found in W/CoFeB bilayer[11]. Consequently, bottom W layer is expected to be efficient for SOT effect.

Based on SOT, the three-terminal magnetic tunnel junction (MTJ), i.e., SOT-MTJ, has been proposed[13]. The SOT-MTJ is quite simple in structure and straight-forward to fabricate[14]. In the SOT-MTJ, the free layer adjoins the HM for switching, and the fixed layer adjoins the other HM for capping. For the writing operation, a charge current is injected into the bottom HM, leading to SOT switching in free layer by the SHE[15]. For the reading operation, tunneling magnetoresistance (TMR) effect is utilized in the vertical MTJ[16]. The separation of writing and reading channels in SOT-MTJ offers additional advantages, such as prolonged endurance, efficient writing operation and high-speed readout[17].Experimental works on SOT-MTJs have been reported[14, 18]. Most recently, a switching current as low as $5.4 \times 10^6$ A/cm$^2$ has been achieved based on SHE in W film[6]. Unfortunately, the TMR in SOT-MTJ is relatively low[16, 18–20]. In SOT-MTJ W(5.2)/Co$_{40}$Fe$_{40}$B$_{20}$(2)/ MgO(1)/Co$_{40}$Fe$_{40}$B$_{20}$(4)/Ta(4)/Ru(5) structure, the TMR is 51%[16], relatively low to that of spin transfer torque MTJ (STT-MTJ)[21], cannot reach the standard of the magnetic random access memory (MRAM).

In this work, we present an *ab initio* investigation on TMR enhancement by symmetrization in SOT-MTJ. We explore the effect on TMR caused by MTJ structure, demonstrating that symmetrical MTJ would enhance TMR due to the resonant tunneling effect.

**METHOD**

Fig. 1 shows the SOT-MTJ atomic structure we consider, which is a two-probe open system with HM as electrodes. The typical structure of the SOT-MTJ is a three-terminal device with a bottom HM layer. The reading operation is based on TMR and performed by flowing a vertical current through the MTJ. As we only discuss the TMR effect in this paper, we built a two-terminal MTJ model as top HM/CoFe/MgO/CoFe/bottom HM. Fivemonolayer (5ML) CoFe is used for both ferromagnetic layers, and 5ML MgO is used as the tunnel barrier layer. Co-O bonds are formed at the CoFe/MgO interface[22]. We use MTJ as the abbreviation of CoFe/MgO/CoFe hetero-junction in the following. The HM/CoFe interfaces have been setup with the crystallographic orientation of HM(001)[110]||CoFe(001)[100] to minimize the lattice mismatch, where we employ the face-centred cubic for HM crystal. In this paper, we choose W as the bottom HM, as giant SHE has been reported in W film[11]. For the top HM, we use W, Ta, Hf, Ru films which are common capping layers[14, 16, 20, 23]. The in-plane lattice constant of the junction is fixed to that of bulk CoFe 2.83°A[24]. SOTMTJ atomic structure was relaxed by *Vienna ab initio simulation package* (VASP)[25–27] until the residual forces on each atom are less than 0.01 eV/°A.

Quantum transport properties were calculated by a state-of-the-art technique based on density functional theory (DFT) combined with the Keldysh non-equilibrium Greens function (NEGF) formalism as implemented in the NanoDCAL package[28, 29]. In the NEGF-DFT transport simulation, the physical quantities are expanded by a linear combination of atomic orbital (LCAO) basis sets at the double-$\zeta$ plus polarization orbital (DZP) level. The Perdew-Burke-Ernzerhof generalized gradient approximation (PBE-GGA) is used as the exchange-correlation potential in all the calculations[30]. The spin-resolved conductance is obtained by the Landauer-Büttiker formula

$$G_\sigma = \frac{e^2}{h} \sum_{k_{||}} T_\sigma(k_{||}, E_F) \tag{1}$$

where $T_\sigma(k_{||}, E_F)$ is the transmission coefficient with spin $\sigma$ at the transverse Bloch wave vector $k_{||} = (k_x, k_y)$ and the Fermi energy $E_F$, $e$ is the electron charge and $h$ is the Planck constant. The spin-resolved transmission coefficient at $E_F$ is calculated by

$$T_\sigma(E_F) = Tr[\Gamma_L(E_F)G^r(E_F)\Gamma_R(E_F)G^a(E_F)])_{\sigma\sigma} \tag{2}$$

where $G^r$ and $G^a$ is the retarded and advanced Greens functions of the system, respectively. $\Gamma_\alpha (\alpha = L, R)$ is the line-width function, which describes the coupling between the $\alpha$ electrode and the scattering region. A 20×20×1 $k$-point mesh was used which was sufficient for the NEGF-DFT self-consistent calculation, and a much denser sampling of 300 × 300 × 1 was used for the calculation the transmission coefficient. The mesh cut-off energy was set to be 3000eV.

After the spin-resolved conductance calculation, we obtained the TMR by the formula

$$TMR = \frac{G_P - G_{AP}}{G_{AP}} \times 100\% \tag{3}$$

where $G_P$ and $G_{AP}$ is the total conductance for the parallel (P) configuration and antiparallel (AP) configuration in SOT-MTJ, respectively. $G_P = G^{\uparrow\uparrow}_P + G^{\downarrow\downarrow}_{AP}$, where $G^{\uparrow\uparrow}_P$ and $G^{\downarrow\downarrow}_P$ is the majority- to majority- spin conductance and minority- to minority- spin conductance, respectively. $G_{AP} = G^{\uparrow\downarrow}_{AP} + G^{\downarrow\uparrow}_{AP}$, where $G^{\uparrow\downarrow}_{AP}$ and $G^{\downarrow\uparrow}_{AP}$ is the majority- to minority- spin conductance and minority- to majority- spin conductance, respectively.

## RESULTS AND DISCUSSIONS

Tab. I shows the spin-resolved conductance and TMR in SOT-MTJs. We built one symmetrical SOT-MTJ, W/MTJ/W, and three asymmetrical SOT-MTJs, including W/MTJ/Ru, W/MTJ/Hf, W/MTJ/Ta. TMR in W/MTJ/W reaches up to around 8500%. All TMRs in asymmetrical MTJs are relatively low. Besides, it can be observed that conductances in AP configurations are similar in all MTJs, while conductances in P vary a lot. The minority-spin conductance in symmetrical MTJ is higher than that in asymmetrical MTJ. For instance, the minority-spin conductance in W/MTJ/W is 20 times higher than that in W/MTJ/Ta. In the following we focus on the comparison between W/MTJ/W and W/MTJ/Ta, as conditions in asymmetrical SOT-MTJs are analogous while Ta is the most common capping material.

To explore the different conductance in W/MTJ/W and W/MTJ/Ta, we firstly studied the electronical structure of W and Ta, as shown in Fig. 2. We found that after the relaxation, the FCC structures distort in z direction and become the body-centered tetragonal structure, so we present the fat band along Γ-Z direction. For CoFe/MgO/CoFe MTJ, $\Delta_5$ state dominates in minority-spin channel in P configuration, as $\Delta_5$ state decays slow in MgO, and minority-spin $\Delta_5$ band crosses with $E_F$ in CoFe[31]. Consequently, we focus on the $\Delta_5$ state, as projected in red color in Fig. 2. For W band, the $\Delta_5$ state passes through the $E_F$, indicating that $\Delta_5$ state could pass through W film. To the contrary, the $\Delta_5$ state in Ta band does not cross with $E_F$, indicating the lack of $\Delta_5$ state in Ta and $\Delta_5$ state would be blocked by Ta film.

Based on the electronical structure analysis, we observe the behavior of density of scattering states (DOSS) at $k_\parallel = (0,0)$ in W/MTJ/W and W/MTJ/Ta, as shown in Fig. 3(a). The behavior in W/MTJ/W is similar to that in W/MTJ/Ta in the beginning W/CoFe/MgO part. However, at the outgoing CoFe/W interface, the DOSS is stable, while at the outgoing CoFe/Ta interface, the DOSS decays rapidly. This phenomenon is attributed to the block effect on $\Delta_5$ caused by Ta film, as demonstrated in Fig. 2. To get a globe understanding of the scattering states behavior, the DOSS over the whole Brillouin zone (BZ) has been researched, as shown in Fig. 3(b). DOSS behaviors in two SOT-MTJs are still similar at the beginning

W/CoFe/MgO part. However, it is remarkable that an overshoot appears at the outgoing MgO/CoFe interface for both SOT-MTJs, and it is more obvious in symmetrical W/MTJ/W. This is the character of resonant tunneling.

To confirm that the resonant tunneling rises in SOT-MTJs, we plot the transmission spectra at $E_F$ in the whole BZ for W/MTJ/W and W/MTJ/Ta as present in Fig. 4. The blue color represents low transmission while the red color represents high transmission. In majority-spin channel in P configurations, the transmission spectra of the majority-spin channels are dominated by the broad peak around the center of BZ, as shown in Fig. 4(a) and 4(c). This is the typical behavior of the coherent tunneling transport for the $\Delta_1$ states[32]. For the minority-spin channel in P configuration shown in Fig. 4(b) and 4(d), some very sharp spikes, called as hot spots, appear at certain $k_\parallel$ points (see red circles), illustrating the resonant tunneling effect. These spectacular spikes occurs due to the symmetrical or nearly symmetrical barriers[33]. In Fig. 4(b) and 4(d), it is prominent that the resonant tunneling in W/MTJ/W is much stronger than that in W/MTJ/Ta. As a result, the conductance in the P configuration, as well as TMR, is strongly enhanced.

In Fig. 5, we prove the IRS by the layer-resolved density of minority-spin $\Delta_5$ states (DOS) for in SOT-MTJs. The red square and blue dot represents W/MT/W and W/MTJ/Ta, respectively. This DOS is at $k_\parallel = (0.48, 0.39)\frac{\pi}{a}$, where the transmission peak reaches up to the maximum in W/MTJ/W. The layer-resolved DOS at this $k_\parallel$ point is symmetrical in W/MTJ/W, and greatly increases at both CoFe/MgO interfaces. Namely, two Co atoms at both MgO sides have sharp DOS in W/MTJ/W, illustrating the IRS occurs, which has significant contribution to the resonant tunneling process, as shown in Fig. 4(b) and Fig. 4(b). This enhances the conductance in P configuration, as well as TMR. However, for asymmetrical W/MTJ/Ta, IRS is weak at the CoFe/MgO interface, explaining the weak resonant tunneling as shown in Fig. 4(b) and Fig. 4(d). Consequently, TMR in asymmetrial W/MTJ/Ta is relatively low.

IRSs produce large resonant tunneling in MTJs if they match identical resonances at the opposite interface. To find out the match condition, we studied the differential density of states (DDOS) at $E_F$. DDOS is the difference of DOS between both Co atoms at opposite MgO interfaces. Green color means the difference is little, red and blue color indicate vary a lot. Fig. 6(a) shows that the DDOS of W/MTJ/W is slight, we attribute the little difference to the perturbation in relaxation. Fig. 6 present that DDOS of W/MTJ/Ta is spectacular. We attribute this phenomenon to the atom positions. When the structure is symmetrical, e.g., W/MTJ/W, the Co atoms at both CoFe/MgO interfaces suffer the same force, the atom positions are almost the same at both sides after relaxation, and the IRSs have perfect match at both MgO sides. When the structure is asymmetrical, e.g., W/MTJ/Ta, the Co atoms at both CoFe/MgO interfaces suffer different force, and the atom positions change at both MgO sides. As a result, the match between both CoFe/MgO interfaces is impaired, leading to the damage of resonant tunneling.

The resonant tunneling is sensitive to the bias and impurities[34]. As shown in Fig. 7(a), the TMR in W/MTJ/W decreases with the voltage, and TMR decline to 6500% at 50mV. This is due to the resonant tunneling damage. The minority-spin conductance at 50mV is one order lower than that at zero bias. However, voltage as low as 1mV is enough for experimental TMR measurement[35], in this condition, the influence caused by voltage is limited and symmetrical structure is beneficial to TMR enhancement. Resonant tunneling is also sensitive to impurities, so perfect CoFe/MgO/CoFe crystallization would optimize the TMR. In addition, we studied how the thickness of bottom layer influences TMR. We built the structure of 1ML W/MTJ/$n$ML W with Ta electrodes, and studied the influence on TMR caused by different bottom W layer thickness. The TMR of 1ML W/MTJ/5ML W is around 11000%, which is used as the normalized standard. All normalized TMRs are present in Fig. 7(b). It can be observed that the TMR only weakly dependent on the thickness. As some works reports the bottom layer thickness influences SOT [10, 20, 36], the TMR is stable for varying bottom layer thickness in SOT-MTJs.

# CONCLUSION

To summarize, we provide a method to enhance the TMR in SOT-MTJ by symmetrization. Compared with W/MTJ/Ta, the TMR in symmetrical W/MTJ/W is optimized thanks to the high minority-spin conductance in P configuration, which is attributed to the character in W electronic structure. Besides, due to the thin MgO layer, resonant tunneling occurs in both SOT-MTJs, which has been evidently illustrated by the overshoot in DOSS at CoFe/MgO interface, and sharp spikes at certain $k_{||}$ points in transmission spectra. We attribute the resonant tunneling to the $\Delta_5$ states in minority-spin channel. Resonant tunneling originates from the match of interfacial resonant states. Compared to the asymmetrical W/MTJ/Ta, the symmetrical W/MTJ/W has higher interfacial resonant states and little difference between atoms at opposite MgO sides. Interfacial resonant states produce large tunneling current in MTJs if they match identical resonances at the opposite barrier interface. As a result, the perfect match in W/MTJ/W significantly contributes to the conductance in P configuration, and enhances the TMR. We also report that low bias and thickness of bottom W layer have limited influence on TMR. Our research provides a feasible approach to enhance TMR in SOT-MTJ, which is promising to solve the low TMR problem in SOT-MTJ and promote the SOT progress.

The authors gratefully acknowledge the National Natural Science Foundation of China (Grant No. 61571023, 61627813), the International Collaboration Project

B16001, and the National Key Technology Program of China 2017ZX01032101 for their financial support of this work.

# Table and Figures

TABLE I. Spin-resolved conductance and TMR in SOT-MTJs. The conductance is in the unit of $10^{-6} \frac{e^2}{h}$.

| model | $G_{PC}^{\uparrow\uparrow}$ | $G_{PC}^{\downarrow\downarrow}$ | $G_{APC}^{\uparrow\downarrow}$ | $G_{APC}^{\downarrow\uparrow}$ | TMR(%) |
|---|---|---|---|---|---|
| W/MTJ/W | 88.4 | 15.0 | 0.6 | 0.6 | 8517 |
| W/MTJ/Ta | 21.2 | 0.7 | 0.4 | 0.9 | 1585 |
| W/MTJ/Ru | 88.1 | 1.8 | 0.7 | 0.7 | 6321 |
| W/MTJ/Hf | 32.8 | 0.6 | 0.2 | 1.8 | 1570 |

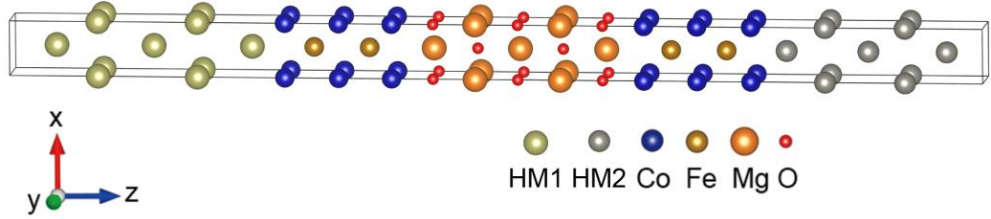

Fig. 1 Atomic structure for the SOT-MTJ. HM1 is the bottom layer, set as W in this paper. HM2 is the capping layer, it can be W, Ta, Ru or Hf.

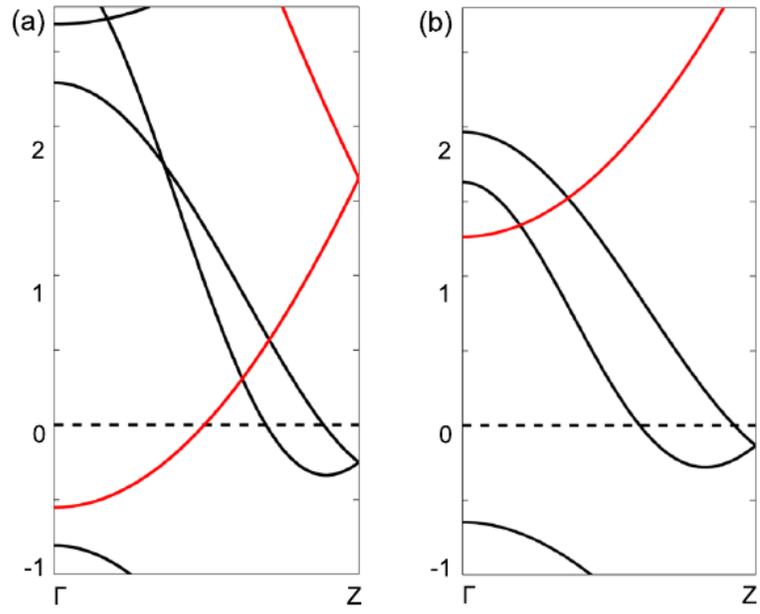

Fig. 2 Electronic structure along Γ-Z direction for (a)W and (b)Ta. $\Delta_5$ states are projected in red color.

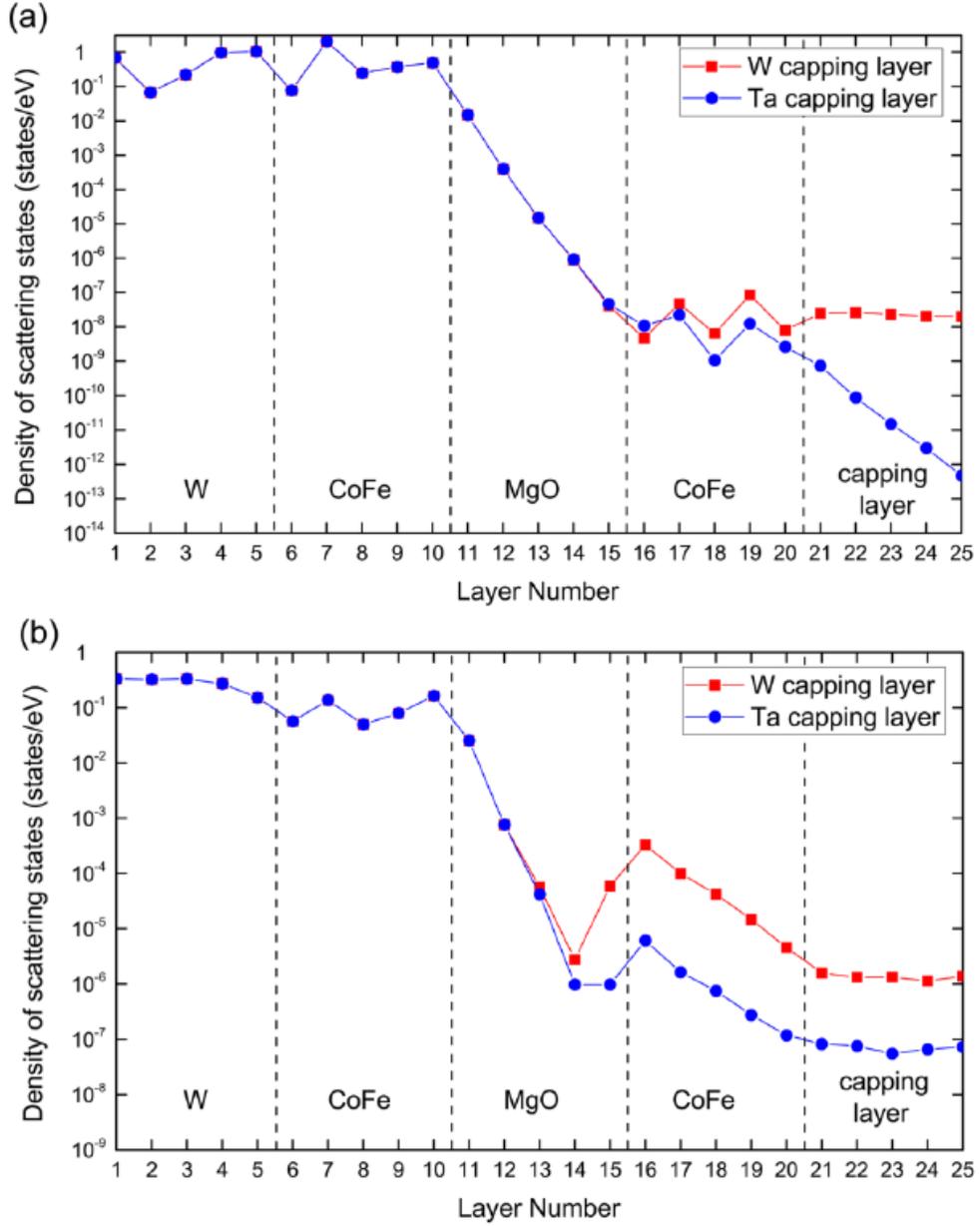

Fig. 3 Density of $\Delta_5$ scattering states (DOSS) for SOT-MTJs at $E_F$. The red square and blue dot represents W/MTJ/W and W/MTJ/Ta, respectively. (a) DOSS at $\mathbf{k}_\parallel=(0, 0)$; (b) DOSS integrated on the whole BZ.

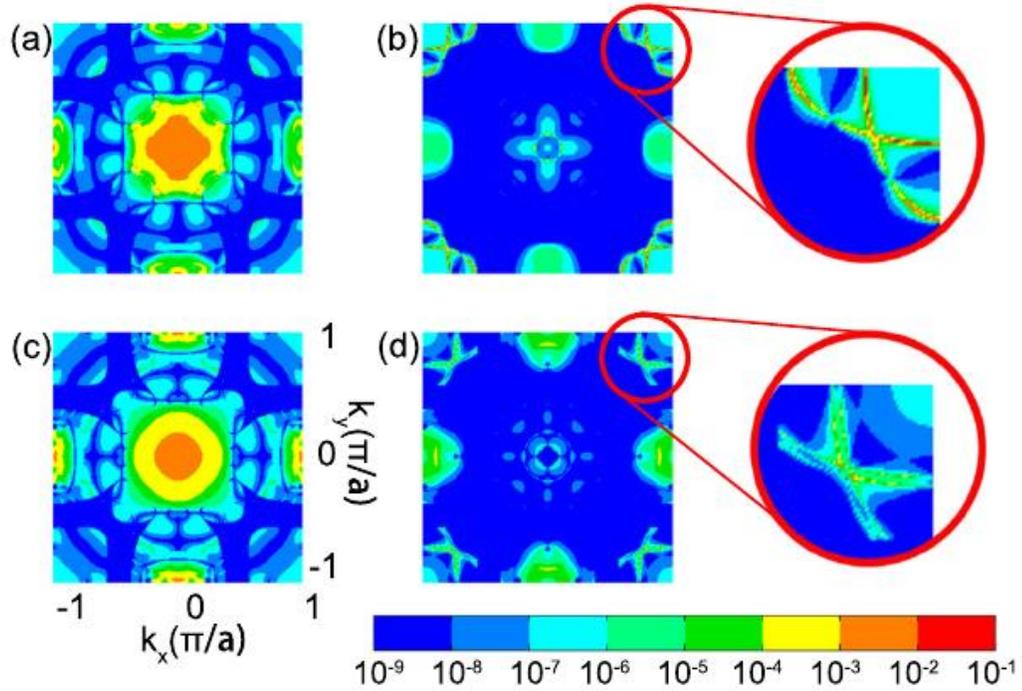

Fig. 4 Spin- and $\mathbf{k}_{\parallel}$-resolved transmission spectra with log scale in parallel condition for (a)majority- to majority- spin channel in W/MTJ/W. (b)minority- to minority- spin channel in W/MTJ/W. (c)majority- to majority- spin channel in W/MTJ/Ta. (d)minority- to minority- spin channel in W/MTJ/Ta. Zoom-in part indicate the resonant tunneling effect in red circles. The color bar indicates the transmission coefficient.

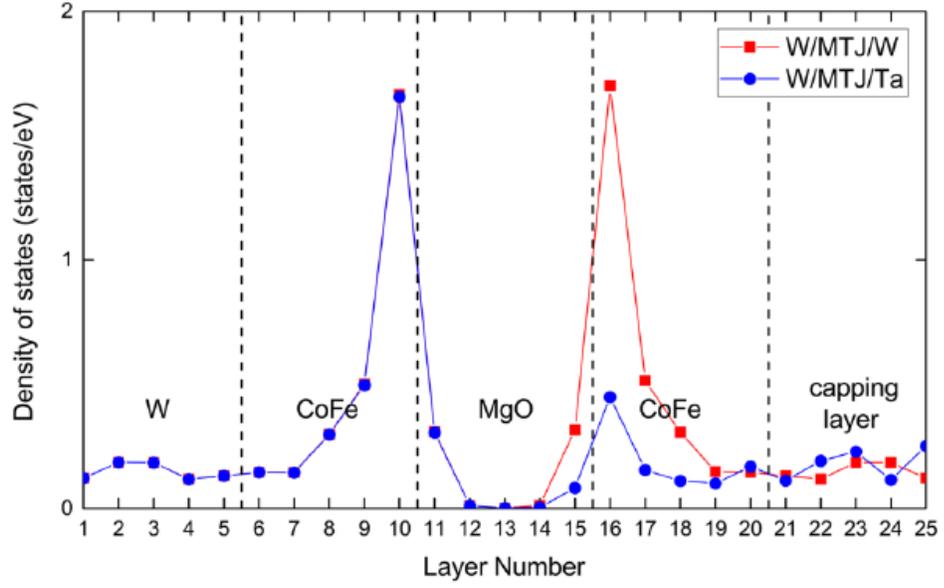

Fig. 5 Layer-resolved density of minority-spin $\Delta_5$ states in SOT-MTJs at $\mathbf{k}_\parallel$=(0.48, 0.39)$\pi$/a at $E_F$. The red square and blue dot represents W/MT/W and W/MTJ/Ta, respectively.

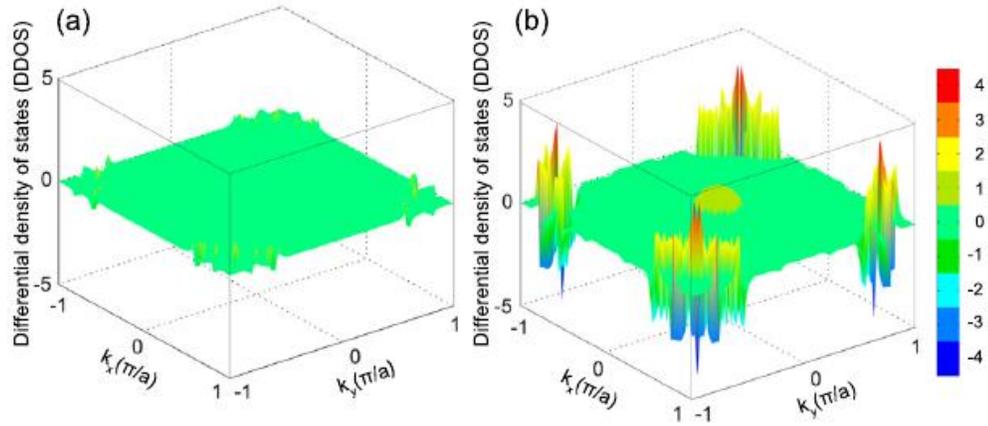

Fig. 6 Differential density of states (DDOS) for Co atoms at both MgO sides at $E_F$ in the whole BZ. (a)W/MTJ/W. (b)W/MTJ/Ta. The olor bar represents the difference.

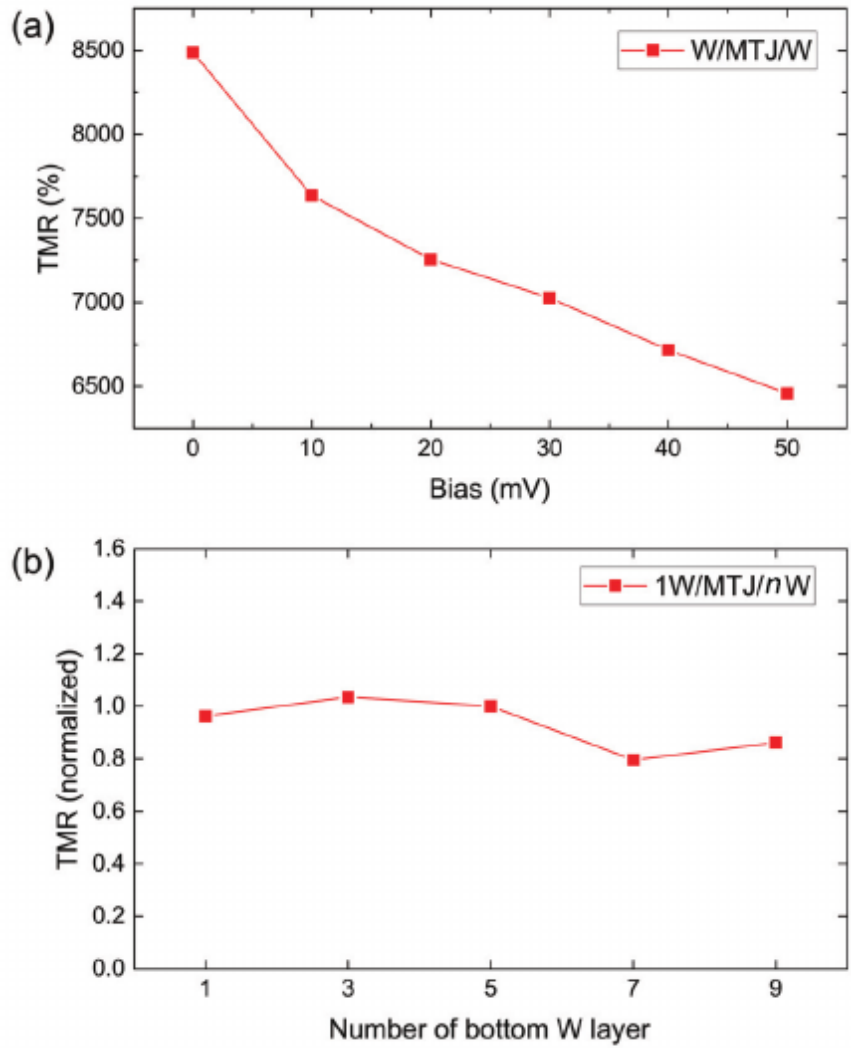

Fig. 7 (a) TMR in W/MTJ/W with bias. (b) TMR in 1ML/MTJ/$n$ML W at zero bias, $n$ is the number of bottom W layer.